\documentclass[12pt]{article}
\usepackage{times}
\begin{document}
\date{}
\author{{S. M. Troshin, N. E.Tyurin}\\
\small\it{Institute for High Energy Physics},\\
\small\it{Protvino, Moscow Region, 142284 Russia}}
\title
{\bf{Diffraction at the LHC -- antishadow scattering?}}
\maketitle
\begin{abstract}
Numerical predictions for the global characteristics of
proton-proton
interactions are given for the LHC energy. Possibilities
for the discovery of the antishadow scattering mode and its
physical implications are discussed.\\
PACS numbers: 12.40.Pp; 13.85.Dz; 13.85.Lg
\end{abstract}

\section*{Introduction}

Soft hadronic interactions observe a time oscillating pattern in
the interest from a high-energy physics community. The peaks of
the interest  coincide as usual with the beginning of the new
machine operation. Nowadays RHIC is preparing for operation and
LHC would start to provide first results in the not too distant
future. Under these circumstances the interest to experimental and
theoretical  studies in this field is increasing.

There are many open problems in  hadron physics at large distances
and their importance has not been overshadowed by the exciting
expectations of the new particles discoveries in the newly opening
energy
 range of the LHC.

The most global characteristic of the hadronic collision is the
total cross--section and the most important problem here is the
nature of the total--cross section rising energy dependence. There
are various approaches which provide total cross-section rising
with energy but the underlying microscopic mechanism leading to
this increase remains obscure. However, the growing understanding
how QCD works at large distances  could finally lead to a final
explanation of this longstanding problem \cite{mart}.

In this connection the TOTEM experiment \cite{vel} approved
recently
 at the LHC could be more valuable
 than  just a tool for checking  numerous model
 predictions and background and luminosity estimates. It could have a
 definite discovery potential and our main
 goal in this note is to discuss one of the such aspects
 related to the possible observation of the antishadow
 scattering mode at the LHC.

\section{When will  asymptotics be seen?}

The answer on the above question currently is model dependent.
There are many model parameterizations for the total
cross-sections  using $\ln^2 s$ dependence for $\sigma_{tot}(s)$.
This implies the saturation of the Froissart--Martin bound and
what is unnatural the presence of the asymptotical
 contributions already at the very moderate energies. On the other
side  the power-like parameterizations of $\sigma_{tot}(s)$
 neglect the Froissart--Martin bound considering it as a matter of the
 distant unknown asymptopia.

It seems that the both  approaches
 are  limited and their limitations reflect
  the real energy range  available for the analysis of the
  experimental data. For example,
  it is not clear whether the  power--like parameterizations respect
  unitarity limit for the partial--wave amplitudes $|f_l(s)|\leq 1$.
   We are keeping in mind here only the accelerator data
  (cosmic ray data will be briefly commented below).

Unitarity is an important principle and the unitarization
procedure of some input power-like ``amplitude'' leads to the
complicated energy dependence of $\sigma_{tot}(s)$ which can be
approximated by the various functions depending on the particular
energy range under consideration. Moreover, unitarity implies the
appearance of the new scattering mode -- antishadow (see
\cite{ech} and references therein). Here we provide numerical
estimates at LHC energies based on the $U$-matrix unitarization method
 \cite{ltkhs} and the particular model for $U$-matrix \cite{chpr}
   and argue that
 antishadow mode could be
revealed already at the LHC energy $\sqrt{s}=14$ TeV.

\section{Antishadow scattering at LHC}
In the impact parameter representation the unitarity equation
written for the elastic scattering amplitude $f(s,b)$
at high energies has the form
\begin{equation}
Im f(s,b)=|f(s,b)|^2+\eta(s,b) \label{unt}
\end{equation}
where the inelastic overlap function $\eta(s,b)$ is the sum of
all inelastic channel contributions.  It can be expressed as
a sum of $n$--particle production cross--sections at the
given impact parameter
\begin{equation}
\eta(s,b)=\sum_n\sigma_n(s,b).
\end{equation}
 Unitarity equation  has the
two solutions for the case of pure imaginary amplitude:
\begin{equation}
f(s,b)=\frac{i}{2}[1\pm \sqrt{1-4\eta(s,b)}].\label{usol}
\end{equation}
Eikonal unitarization with pure imaginary eikonal corresponds to the
choice of the particular
solution with sign minus.

In the $U$--matrix approach
the form of the elastic scattering amplitude in the
impact parameter representation
is the following:
\begin{equation}
f(s,b)=\frac{U(s,b)}{1-iU(s,b)}. \label{um}
\end{equation}
 $U(s,b)$ is the generalized reaction matrix, which is considered as an
input dynamical quantity similar to eikonal function.

Inelastic overlap function
is connected with $U(s,b)$ by the relation
\begin{equation}
\eta(s,b)=\frac{Im U(s,b)}{|1-iU(s,b)|^{2}}\label{uf}.
\end{equation}

 Construction of  particular models in the framework of the $U$--matrix
approach proceeds the standard steps, i.e. the basic dynamics as well as
the notions on hadron structure
are used  to obtain a particular form for the $U$--matrix.

However, the two unitarization schemes ($U$--matrix and eikonal)
 lead to
different predictions for
 the inelastic cross--sections and for the ratio of elastic to total
cross-section. This ratio in the $U$--matrix unitarization scheme
reaches its maximal possible value at $s\rightarrow \infty$, i.e.
\begin{equation}
\frac{\sigma_{el}(s)}{\sigma_{tot}(s)}\rightarrow 1,
\end{equation}
which reflects in fact that the bound for the partial--wave
 amplitude in the $U$--matrix
approach is $|f(s,b)|\leq 1$ while  the bound for the case of
imaginary eikonal is (black disk limit): $|f(s,b)|\leq 1/2$.

When the amplitude exceeds the black disk limit (in central
collisions at high energies) then the scattering at such impact
parameters turns out to be of an  antishadow nature.
 In this antishadow scattering mode
 the elastic amplitude increases with decrease of the inelastic
 channels contribution.

The shadow scattering mode is  considered usually as  the only possible
one. But the two solutions of the unitarity
 equation have an equal meaning and the antishadow scattering mode could also
appear in the central collisions first as the energy becomes higher.
The both scattering modes are realized in a natural way under
the $U$--matrix unitarization despite the two modes are described by the two
 different solutions of unitarity.

 Appearance of the antishadow scattering mode
is consistent with the basic idea that the particle production
is the driving force for elastic scattering. Indeed, the imaginary
part of the generalized reaction matrix is the sum of inelastic channel
 contributions:
\begin{equation}
Im U(s,b)=\sum_n \bar{U}_n(s,b),\label{vvv}
\end{equation}
where $n$ runs over all inelastic states and
\begin{equation}
\bar{U}_n(s,b)=\int d\Gamma_n |U_n(s,b,\{\xi_n\}|^2
\end{equation}
and $d\Gamma_n$ is the $n$--particle element of the phase space
volume.
The functions $U_n(s,b,\{\xi_n\})$ are determined by the dynamics
 of $2\rightarrow n$ processes. Thus, the quantity $ImU(s,b)$ itself
 is a shadow of the inelastic processes.
However, unitarity leads to  self--damping of the inelastic
channels \cite{bbl} and increase of the function $ImU(s,b)$ results in
decrease
 of the inelastic overlap function $\eta(s,b)$ in accord with Eq. (\ref{uf})
  when $ImU(s,b)$
  exceeds unity.

Let us consider the transition to the antishadow scattering mode
 \cite{phl}. With
conventional parameterizations of the $U$--matrix
 the inelastic overlap function increases with energies
at modest values of $s$. It reaches its maximum value $\eta(s,b=0)=1/4$ at some
energy $s=s_0$ and beyond this energy the  antishadow
scattering mode appears at small values of $b$. The region of energies and
impact parameters corresponding
to the antishadow scattering mode is determined by the conditions
$Im f(s,b)> 1/2$ and $\eta(s,b)< 1/4$.
The quantitative analysis of the experimental data
 \cite{pras} gives the threshold value: $\sqrt{s_0}\simeq 2$ TeV.

Thus, the function $\eta(s,b)$ becomes peripheral when energy is increasing.
At such energies the inelastic overlap function reaches its maximum
 value at $b=R(s)$ where $R(s)$ is the interaction radius.
So, beyond the transition threshold there are two regions in impact
 parameter space: the central region
of antishadow scattering at $b< R(s)$ and the peripheral region
of shadow scattering at $b> R(s)$.

 The region of the LHC energies is the one where antishadow scattering
 mode is to be presented. It will be demonstrated in the next
 section that this mode can be revealed directly measuring
 $\sigma_{el}(s)$ and $\sigma_{tot}(s)$ and not only through the
 analysis of impact parameter distributions.
 \section{Estimates and transition to asymptotics}
We use chiral quark model for the $U$--matrix \cite{chpr}.  The
function $U(s,b)$  is chosen in the model as a product of the
averaged quark amplitudes \begin{equation} U(s,b) =
\prod^{N}_{Q=1} \langle f_Q(s,b)\rangle \end{equation} in
accordance  with assumed quasi-independent  nature  of the valence
quark scattering in some effective field. The essential point here
is the rise with energy of the number of the scatterers  like
$\sqrt{s}$ (cf. \cite{chpr}). The $b$--dependence of the function
$\langle f_Q \rangle$ is related to
 the quark formfactor $F_Q(q)$ and has a simple form $\langle
f_Q(b)\rangle\propto\exp(-m_Qb/\xi )$, i.e. the valence quarks in
the model have a complicated structure with quark matter
distribution approximated by the function $\langle f_Q(b)\rangle$.

The generalized
reaction matrix (in a pure imaginary case) gets
the following  form
\begin{equation} U(s,b) = ig\left [1+\alpha
\frac{\sqrt{s}}{m_Q}\right]^N \exp(-Mb/\xi ), \label{x}
\end{equation} where $M =\sum^N_{Q=1}m_Q$.
Here $m_Q$ is the mass of constituent quark, which is taken to be
$0.35$ $GeV$, $N$ is the total number of valence quarks in the
colliding hadrons, i.e. $N=6$ for $pp$--scattering. The values for
the other parameters were obtained in \cite{pras}:
 $g=0.24$, $\xi=2.5$, $\alpha=0.56\cdot 10^{-4}$.
These parameters were adjusted to the experimental data on the
total cross--sections in the range up to the Tevatron energy. With
such small number of  free parameters the model is in a rather
good agreement with the data \cite{pras}.

For the LHC energy $\sqrt{s}= 14$ $TeV$ the model gives
\begin{equation}\label{s}
 \sigma_{tot}\simeq 230\; \mbox{mb}
\end{equation}
and
\begin{equation}\label{r}
\sigma_{el}/\sigma_{tot}\simeq 0.67.
\end{equation}
 Thus, the antishadow scattering mode could be discovered
at the LHC by measuring $\sigma_{el}/\sigma_{tot}$ ratio which is
greater than the black disc value $1/2$.

However, the LHC energy is not in the asymptotic region yet;
the total, elastic and inelastic cross-sections behave like
\begin{equation}\label{tot}
  \sigma_{tot,el}\propto \ln^2\left[g\left(1+\alpha
\frac{\sqrt{s}}{m_Q}\right)^N\right],\;
\end{equation}
\begin{equation}\label{ine}
 \sigma_{inel}\propto \ln\left[g\left(1+\alpha
\frac{\sqrt{s}}{m_Q}\right)^N\right].
\end{equation}
True asymptotical regime
\begin{equation}\label{tota}
  \sigma_{tot,el}\propto \ln^2 s,\;\;
 \sigma_{inel}\propto \ln s
\end{equation}
is expected at $\sqrt{s}> 100$ $TeV$.

Another predictions of the chiral quark model is decreasing energy
dependence of the the cross-section of the inelastic diffraction
at $s>s_0$. Decrease of diffractive production cross--section at high energies
($s>s_0$) is due to the  fact  that $\eta  (s,b)$
becomes peripheral at  $s > s_0$  and  the whole  picture  corresponds  to
the antishadow scattering at $b < R(s)$ and to the shadow scattering at
$b>R(s)$ where $R(s)$ is the interaction radius:
\begin{equation}
\frac{d\sigma_{diff}}{dM^2_X}\simeq
\frac{8\pi  g^*\xi ^2}{M_X^2} \eta(s,0).
\end{equation}
The parameter $g^*<1$ is the probability of the excitation of a
constituent quark during interaction.
Diffractive production cross--section  has  familiar
$1/M_X^2$  dependence     which
is related in  this model to the geometrical size  of excited  constituent
quark.

At the LHC energy $\sqrt{s}=14$ $TeV$ the value of the single
diffractive inelastic cross-sections is limited by the value
\begin{equation}\label{ind}
  \sigma _{diff}(s)\leq 2.4\;\mbox{mb}.
\end{equation}

The above predicted values for the global characteristics of
$pp$ -- interactions at the LHC differ from the most common predictions of
the other models. First of all total cross--section is predicted
to be twice as much of  the standard predictions in the range 95-120
mb \cite{vels}
 and it  also overshoots the existing cosmic ray data. However,
 extracting proton--proton cross sections from cosmic ray
 experiments is model dependent and  far from straightforward
 (see, e.g. \cite{bl} and references therein). Those experiments
 measure the attenuation lengths of showers initiated  by the cosmic
 particles in the atmosphere and are sensitive to the model
 dependent parameter called inelasticity. So the disagreement of the
  particular model with
 the cosmic ray measurements means that the data should be
 recalculated in the framework of this model and in addition
assumptions on the energy dependence of inelasticity should
be involved also.

\section{Discussions and conclusion}
The main goal of this note is to point out that the antishadow
scattering mode at the LHC can be  detected measuring elastic to total
cross section ratio which is predicted to be greater than the
black disc limit $1/2$. The considered model estimates also the
total cross section values significantly higher than the values
conventional parameterizations provide.

 The studies of soft interactions at the LHC energies can
lead to the discoveries of fundamental importance. The genesis
 of hadron scattering with rising energy
 can be described as  transition from the grey to black disc
  and eventually
to black ring with the antishadow scattering mode in the center.
It is worth noting that the appearance of the antishadow
scattering mode at the LHC  implies a somewhat unusual
scattering picture. At high energies the proton should be
represented as a very loosely bounded composite system and it
appears that this system has a high probability to reinstate
itself only in the central collisions where all of its parts
participate in the coherent interactions. Therefore the central
collisions are mostly responsible for elastic processes while the
peripheral ones where only few parts of weekly bounded protons are
involved result mainly in the production of  secondary particles. This
leads to the peripheral impact parameter profile of the inelastic
overlap function. The above picture would imply interesting
consequences for the multiplicities in hadronic collisions,
 i.e. up to the threshold energy $s_0$ the picture will correspond
to the  fragmentation concept \cite{ben} which supposes  larger
multiplicity for the higher value of momentum
 transfer. The increase of the mean multiplicity in
hadron interactions with $t$ \cite{tro} is in agreement with the
hadronic  experimental data. However, when the energy  becomes
greater than $s_0$ and antishadow mode develops,
 momentum transfer dependence of
multiplicity would change.
Loosely speaking  the picture described above correspond to the
scattering of extended objects at lower energies
 and transition to the scattering of weakly
bounded  systems at higher energies. This picture has an
illustrative value and is in general compliance with asymptotic
freedom of QCD and parton picture.

Finally, we would like to note that the numerical predictions
depend on the particular choice of the model for the $U$-matrix,
but appearance of the antishadow scattering mode is an
inherent feature of the considered approach.
\section*{Acknowledgements}
 Authors are grateful to W.~Kienzle,  A.~Krisch,
 W.~Lorenzon and V.~Roinishvili for the interesting discussions.
 This work was supported in part by the RFBR Grant No. 99-02-17995.


\small
\begin{thebibliography}{99}
\bibitem{mart}
A. Martin,  CERN-TH.7284/94 Preprint, 1994.
\bibitem{vel}
G. Matthiae,
In ``Future Physics and Accelerators'',
 Edited by M.
    Chaichian, K. Huitu, R. Orava,
     World Scientific, 1995, 245.
\bibitem{ech}
S. M. Troshin and N. E. Tyurin,
Phys. Part. Nucl. 30 (1999) 550.
\bibitem{ltkhs}
A. A. Logunov, V. I. Savrin, N. E. Tyurin and O. A. Khrustalev,
 Teor. Mat. Fiz.
\bf 6 \rm (1971) 157;
\bibitem{chpr} S.M. Troshin and N.E.
Tyurin, Nuovo Cim.  \bf  106A \rm  (1993) 327; Proc. of the Vth
Blois
 Workshop on Elastic and Diffractive Scattering, Providence, Rhode
Island, June 1993, p. 387; Phys. Rev. \bf D49 \rm (1994) 4427; Z.
Phys. C\bf 64 \rm (1994) 311.
\bibitem{bbl}
M. Baker and R. Blankenbecler, Phys. Rev. \bf 128 \rm (1962) 415.
\bibitem{phl}
S. M. Troshin and N. E. Tyurin, Phys. Lett. \bf B 316  \rm (1993)
175.
\bibitem{pras}
 P. M. Nadolsky, S. M. Troshin and N. E. Tyurin, Z. Phys.
C \bf  69 \rm (1995)   131 .
\bibitem{bl} M. M. Block, F. Halzen
and T. Stanev, hep-ph/9908222 Preprint, 1999.
\bibitem{vels}
J. Velasco , J. Perez-Peraza, A. Gallegos-Cruz, M.
Alvarez-Madrigal, A. Faus-Golfe, A. Sanchez-Hertz,
  hep-ph/9910484 Preprint, 1999.
\bibitem{ben}
J. Benecke, T.~T.~Chou, C. N. Yang and E. Yen, Phys. Rev. \bf 188
\rm (1969) 2159; T.~T.~Chou and C. N. Yang, Phys. Rev. D \bf 50
\rm (1994) 590.
\bibitem{tro}
S. M. Troshin, Sov. J. Nucl. Phys. \bf 25 \rm (1977) 472.

\end{thebibliography}
\end{document}